\documentclass[12pt]{article}
\usepackage[margin=1in]{geometry}
\usepackage{graphicx}
\usepackage{caption}
\usepackage{subcaption}
\usepackage{float}
\usepackage{amssymb, amsmath, amsthm, mathrsfs}
\allowdisplaybreaks
\usepackage{hyperref}
\usepackage{natbib}
\usepackage{tocloft}
\usepackage{soul, multirow}

\usepackage[T1]{fontenc}
\usepackage{tikz}
\newcommand\tick[1][]{%
    \tikz\draw[thick, line cap=round, x=1ex, y=1ex, #1]
    (0,0) -- ++(90:0.95)
    (0,0) -- ++(-30:0.9)
    (0,0) -- ++(-150:0.9);\ignorespacesafterend}

\usepackage{xcolor}

\theoremstyle{definition}
\newtheorem{definition}{Definition}

\newcommand{\bm}[1]{\boldsymbol{#1}}

\title{Automatic selection of hyper-parameters via the use of softened profile likelihood}
\date{\today}
\author{Gengyang Chen\footnote{Python code for this project is available at \url{https://github.com/GyangC/automatic_selection}.}, Mu Zhu\footnote{This research is supported by grant RGPIN-2023-03337 from the Natural
Sciences and Engineering Research Council (NSERC) of Canada.}\\[3mm]Department of Statistics and Actuarial Science\\University of Waterloo\\Waterloo, Ontario, Canada N2L 3G1}

\begin{document}
\maketitle

\begin{abstract}
We extend a heuristic method for automatic dimensionality selection, which maximizes a profile likelihood to identify ``elbows'' in scree plots. Our extension enables researchers to make automatic choices of multiple hyper-parameters simultaneously. To facilitate our extension to multi-dimensions, we propose a ``softened'' profile likelihood. We present two distinct parameterizations of our solution and demonstrate our approach on elastic nets, support vector machines, and neural networks. We also report a small simulation study to investigate violations to an assumption we make, and briefly discuss applications of our method to other data-analytic tasks than hyper-parameter selection.
\end{abstract}

\newpage
\tableofcontents
\newpage

\section{Introduction}
\label{sec:intro}

\begin{figure}[hpt]
\centering
    \begin{subfigure}[t]{0.495\textwidth}
        \centering
    \includegraphics[width=\textwidth]{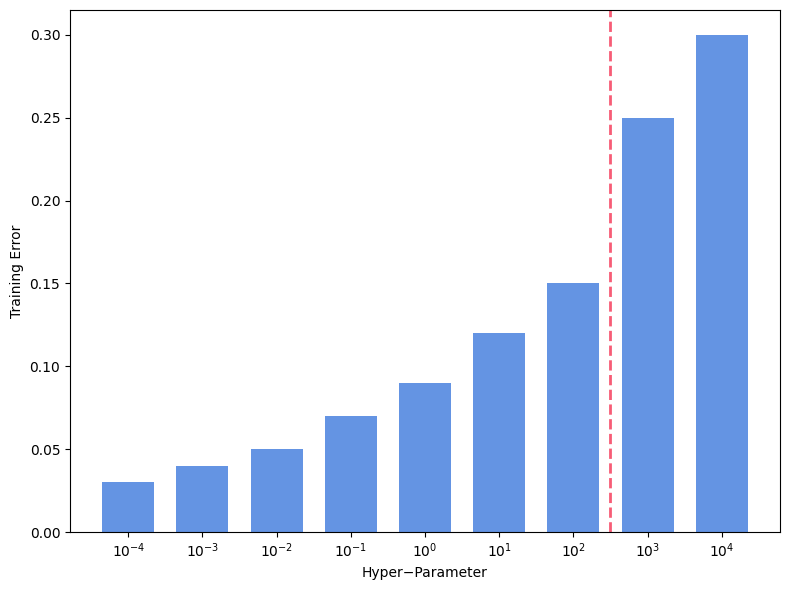}
    \caption{\label{fig:scree}One hyper-parameter.}
    \end{subfigure}
    \begin{subfigure}[t]{0.495\textwidth}
        \centering
    \includegraphics[width=\textwidth]{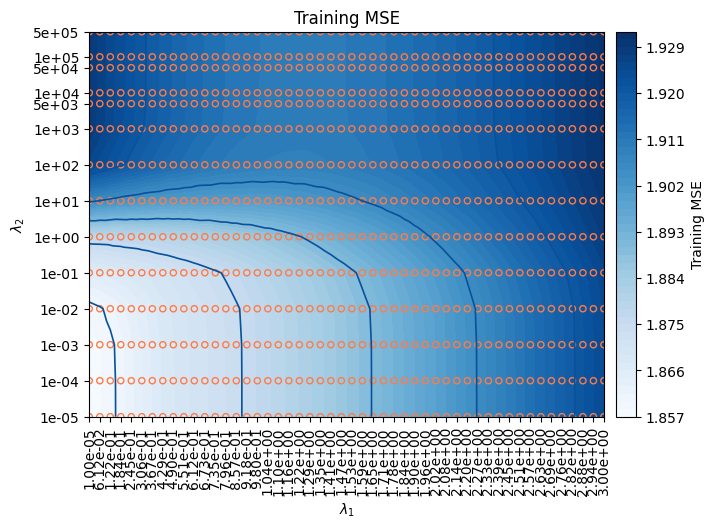}
    \caption{\label{fig:ENcontour}Two hyper-parameters.
    }
    \end{subfigure}
\caption{\label{fig:open_example}Examples of hyper-parameters affecting model performance.}
\end{figure}

Researchers who use computational tools often look at a so-called ``scree plot'' to make choices about certain hyper-parameters in their numeric procedure. Figure~\ref{fig:scree} shows a prototypical scenario: A team of researchers may be training a model while using a certain hyper-parameter to penalize its complexity. When the penalty is small, relatively low training errors can be achieved, but the researchers understand the model is likely overfitting the training data. As the penalty is increased, the error grows as well. The growth is slow and steady at first, but becomes much steeper ``at some point''---here, marked by a dashed vertical line in Figure~\ref{fig:scree}. It is this ``turning point'' that is critical to the researchers. For example, they may conclude that the reduction in model complexity ``up to this point'' will more than compensate for the slight increase in training error, but that any continued reduction ``beyond this point'' will no longer be justifiable against the sharp increase in error. 

\citet{zhu2006} provided a heuristic method to automate such decisions. The need for automation is most prominent when running repeated simulations. Without an automatic procedure, researchers would have to stop each simulation in the middle to inspect the ``scree plot'' before an appropriate hyper-parameter can be chosen and the simulation can be continued. Over the past two decades, their method has garnered reasonably wide attention and adoption by researchers from different fields, such as 
bioinformatics~\citep[e.g.,][]{fogel_young2007bioinf},
demographics~\citep[e.g.,][]{Zhong2021actuarial},
ecology~\citep[e.g.,][]{hamady2010ecology},
genetics~\citep[e.g.,][]{lai2009genetics},
geoscience~\citep[e.g.,][]{Honarkhah2010487}, 
and
physics~\citep[e.g.,][]{chang2022physics}. 
It has also become a frequently used tool in 
network data analysis~\citep[e.g.,][]{priebe2017ieee_network},
the study of random graphs~\citep[e.g.,][]{priebe2017jcgs}, 
and
spectral clustering~\citep[e.g.,][]{priebe2019pnas}.

Although it can be used generally for hyper-parameter selection, the heuristic method of \citet{zhu2006} was originally presented for the purpose of determining the number of principal components, and not applicable to problems with more than one hyper-parameter; we will elaborate on this limitation later in Section~\ref{sec:generalization} to make it more clear. 

On the other hand, it is now common for many statistical and computational problems to have more than one hyper-parameter. A highly-cited example in statistics is the elastic net \citep{enet}:
\begin{equation}
\underset{\bm{\beta}}{\min} \ 
\|\bm{y}-\bm{X}\bm{\beta}\|^2_{\ell_2} 
+ \lambda_1 \|\bm{\beta}\|_{\ell_1}
+ \lambda_2 \|\bm{\beta}\|^2_{\ell_2}
\label{eq:EN}
\end{equation}
with two hyper-parameters $\lambda_1, \lambda_2 > 0$. 
Figure~\ref{fig:ENcontour} shows an example of fitting an elastic net model~\eqref{eq:EN} to 
a dataset about forest fires in Portugal. 
We will return to this example in Section~\ref{sec:example_EN} with more details, but the decision faced here by the data analyst can already be appreciated: 
As both hyper-parameters increase, the mean squared error (MSE) increases, slowly at first, but ``at some point'' the increase starts to accelerate in both directions.

The method of \citet{zhu2006} can be used to determine the ``turning point'' in Figure~\ref{fig:scree}, but not the one in Figure~\ref{fig:ENcontour}. In this article, we extend their work to overcome this limitation. 

\section{Methodology}
\label{sec:main}

In a nutshell, our primary objective is to develop a method for identifying a boundary in the space of hyper-parameters at which improvements in model performance can be considered to have plateaued. Specifically, we aim to estimate a boundary curve---say, $g(\bm{u}; \boldsymbol{\omega}) = 0$---and a representative point on it, from a dataset of $n$ performance evaluations $\{(\bm{u}_i, z_i)\}_{i=1}^n$, where each $\bm{u}_i$ denotes a specific hyper-parameter configuration and $z_i$, the corresponding value of model performance. Our methodology involves constructing a model for $z_i$ and estimating the boundary curve by maximizing a profile log-likelihood function. We also provide a principled procedure for selecting a representative point on this boundary, and recommend a simple transformation of $\bm{u}_i$ in practice to improve numerical stability during optimization.

\subsection{Brief review}
\label{sec:review}

We start with a brief summary of \citet{zhu2006}. Suppose $u_1, u_2, \dots, u_n$ are a set of candidate values for a hyper-parameter, and $z_1, z_2, \dots, z_n$ are the corresponding values of an evaluation metric under consideration. 
Their idea is to model each $z_i$ as an independent observation from either of two different distributions depending on whether $u_i$ is above or below a certain threshold $\omega$, i.e.,
\[
z_i \sim
\begin{cases}
f(z; \bm{\theta}_1), & \text{if} \quad u_i \leq \omega; \\[1mm]
f(z; \bm{\theta}_2), & \text{if} \quad u_i > \omega. 
\end{cases}
\]
The log-likelihood is then given by
\begin{equation*}
    \ell(\omega, \boldsymbol{\theta}_1, \boldsymbol{\theta}_2) = 
    \sum_{u_i \leq \omega} \log f(z_i;\boldsymbol{\theta}_1) + 
    \sum_{u_i > \omega} \log f(z_i;\boldsymbol{\theta}_2),
\end{equation*}
and the optimal threshold is chosen by maximizing the profile log-likelihood,
\begin{equation}
    \ell_{\omega}(\omega) = 
    \sum_{u_i \leq \omega} \log f(z_i;\hat{\boldsymbol{\theta}}_1(\omega)) + 
    \sum_{u_i > \omega} \log f(z_i;\hat{\boldsymbol{\theta}}_2(\omega)),
    \label{eq:objective_2006}
\end{equation}
where 
$\hat{\boldsymbol{\theta}}_j(\omega)$ is the maximum likelihood estimator (MLE) of $\bm{\theta}_j$ for any given $\omega$, with $j\in\{1,2\}$. 

\subsection{Generalization}
\label{sec:generalization}

If we have more than one hyper-parameter, then $\bm{u}_1, \bm{u}_2, \dots, \bm{u}_n \in \mathbb{R}^m$ with $m>1$. Clearly, this multidimensional space can no longer be separated into two regions by a simple threshold. That is why the method of \citet{zhu2006} no longer applies. Instead, we will need a ``boundary function'', 
say, 
$g(\bm{u}; \bm{\omega})$,
parameterized here by $\bm{\omega}$; it will partition the space into
\[
\{\bm{u}\in\mathbb{R}^m: g(\bm{u}; \bm{\omega}) \leq 0\} \quad\text{and}\quad
\{\bm{u}\in\mathbb{R}^m: g(\bm{u}; \bm{\omega}) > 0\},
\]
depending on which side of the boundary $\bm{u}$ lies. Conceptually, we can simply obtain the optimal boundary by maximizing a similar profile log-likelihood,
\begin{equation}
    \ell_{\bm{\omega}}(\bm{\omega}) = 
    \sum_{g(\bm{u}_i; \bm{\omega}) \leq 0} \log f(z_i;\hat{\boldsymbol{\theta}}_1(\bm{\omega})) + 
    \sum_{g(\bm{u}_i; \bm{\omega}) > 0} \log f(z_i;\hat{\boldsymbol{\theta}}_2(\bm{\omega})),
    \label{eq:objective_naive}
\end{equation}
completely analogous to Eq.~\eqref{eq:objective_2006}. In particular, Eq.~\eqref{eq:objective_2006} is easily seen to be a special case of Eq.~\eqref{eq:objective_naive}, with $g(u;\omega)=u-\omega$ and $u, \omega \in \mathbb{R}$. However, because Eq.~\eqref{eq:objective_naive} is not a continuous function of $\bm{\omega}$, some further modification is needed (details in Section~\ref{sec:softening} below) to make the optimization problem practical to solve. 

\subsection{Softening}
\label{sec:softening}

Since \citet{zhu2006} only dealt with the one-dimensional case of $u_i \in \mathbb{R}$, a direct grid search---typically just over the entire set of candidate values $\{u_1, u_2, \dots, u_n\}$---was sufficient in practice for them to maximize Eq.~\eqref{eq:objective_2006} over $\omega$. For $\bm{u}_i \in \mathbb{R}^m$ with $m>1$, such an approach is no longer applicable, and we will need to use numerical optimization tools to maximize Eq.~\eqref{eq:objective_naive} over $\bm{\omega}$, but the fact that Eq.~\eqref{eq:objective_naive} is not continuous in $\bm{\omega}$ poses an extra challenge. 

Our main idea here is to turn Eq.~\eqref{eq:objective_naive} into a continuous function of $\bm{\omega}$ by replacing the ``hard'' assignment of each $\bm{u}_i$ to one side of the boundary, i.e.,
\[
\text{either}\quad 
g(\bm{u}_i; \bm{\omega}) \leq 0 \quad\text{or}\quad
g(\bm{u}_i; \bm{\omega}) > 0, 
\]
with a ``soft'' weight based on the sigmoid function,
\begin{equation}
s_i(\boldsymbol{\omega}) := \frac{1}{1 + \exp(-g(\bm{u}_i; \boldsymbol{\omega}))},
\label{eq:weight}
\end{equation}
which is close to zero if $\bm{u}_i$ is on one side of the boundary $g(\bm{u}_i; \bm{\omega}) \leq 0$, and close to one if it is on the other side $g(\bm{u}_i; \bm{\omega}) > 0$.

More specifically, the ``softened'' profile log-likelihood becomes 
\begin{equation}
    \ell^{\text{soft}}_{\boldsymbol{\omega}}(\boldsymbol{\omega}) = 
    {\sum_{i=1}^n} (1-s_i(\bm{\omega})) \log f\big(z_i;\hat{\boldsymbol{\theta}}^{\text{soft}}_1(\boldsymbol{\omega})\big) +  
    {\sum_{i=1}^n} s_i(\bm{\omega}) \log f\big(z_i;\hat{\boldsymbol{\theta}}^{\text{soft}}_2(\boldsymbol{\omega})\big),
    \label{eq:objective_soft}
\end{equation}
where
\begin{equation}
\left(
\hat{\bm{\theta}}^{\text{soft}}_1(\bm{\omega}), 
\hat{\bm{\theta}}^{\text{soft}}_2(\bm{\omega})
\right) = 
\underset{\left(\bm{\theta}_1,\bm{\theta}_2\right)}{\arg\max} \ 
\left\{
{\sum_{i=1}^n} (1-s_i(\bm{\omega})) \log f(z_i; \bm{\theta}_1) +
{\sum_{i=1}^n} s_i(\bm{\omega}) \log  f(z_i; \bm{\theta}_2)
\right\},
\label{eq:thetaMLE}
\end{equation}
are ``softened'' MLEs given $\bm{\omega}$. 
One can easily see that both the original profile log-likelihood \eqref{eq:objective_naive} and the ``usual'' MLEs given $\bm{\omega}$,
\begin{equation*}
\left(
\hat{\bm{\theta}}_1(\bm{\omega}),
\hat{\bm{\theta}}_2(\bm{\omega})
\right) = 
\underset{\left(\bm{\theta}_1,\bm{\theta}_2\right)}{\arg\max} \ 
\left\{
\sum_{g(\bm{u}_i;\bm{\omega}) \leq 0} \log f(z_i; \bm{\theta}_1) + 
\sum_{g(\bm{u}_i;\bm{\omega}) > 0} \log f(z_i; \bm{\theta}_2)
\right\},
\end{equation*}
are simply special cases of \eqref{eq:objective_soft} and \eqref{eq:thetaMLE}, respectively, with 
\[
s_i(\bm{\omega}) = 
\begin{cases}
0, & g(\bm{u}_i;\bm{\omega}) \leq 0;\\
1, & g(\bm{u}_i;\bm{\omega}) > 0
\end{cases}
\]
being a step function, which is what caused the discontinuity problem mentioned earlier at the end of Section~\ref{sec:generalization}. 

\subsection{Choice of $f$}
\label{sec:f_choice}

\citet{zhu2006} assumed 
\begin{equation}
f(z; \bm{\theta}_j) := f(z;\mu_j,\sigma^2) = \frac{1}{\sqrt{2\pi \sigma^2}} \exp\left\{-\frac{(z - \mu_j)^2}{2\sigma^2}\right\}, \quad j \in \{1,2\}. 
\label{eq:f_choice}
\end{equation}
That is, the two different distributions were assumed to be Gaussian with a different mean but a common variance. 

This was the most arbitrary assumption in their work, but also what allowed their automatic procedure to produce threshold choices in one dimension that best ``matched'' what an ``average'' data scientist would otherwise choose by visual inspection. Therefore, we shall keep this assumption in this article. Examples later in Section~\ref{sec:examples} will confirm that doing so does indeed allow us to produce boundaries that \emph{continue} to ``match'' what an ``average'' data scientist would otherwise sketch out in the space of $\bm{u}_1, \bm{u}_2, \dots, \bm{u}_n$ by visual inspection. We also investigate violations to this assumption with a small simulation study in Appendix~\ref{app:sim}.

Straightforward derivations omitted here but provided in Appendix~\ref{app:A} will now allow us to conclude that, under assumption \eqref{eq:f_choice}, the ``softened'' profile log-likelihood \eqref{eq:objective_soft} can be further reduced to
\begin{equation}
\ell^{\text{soft}}_{\bm{\omega}}(\bm{\omega}) = -\frac{n}{2} \log(2\pi\hat{\sigma}^2(\boldsymbol{\omega})) - \frac{n}{2}, \label{eq:objective_simplified}
\end{equation}
where
\[
\hat{\sigma}^2(\boldsymbol{\omega}) = \frac{\mathrm{RSS}_1(\boldsymbol{\omega}) + \mathrm{RSS}_2(\boldsymbol{\omega})}{n},
\]
with
\begin{equation*}
    \mathrm{RSS}_1(\boldsymbol{\omega}) = \sum_{i=1}^n (1-s_i(\boldsymbol{\omega})) (z_i - \hat{\mu}_1(\boldsymbol{\omega}))^2, \quad
    \mathrm{RSS}_2(\boldsymbol{\omega}) = \sum_{i=1}^n s_i(\boldsymbol{\omega}) (z_i - \hat{\mu}_2(\boldsymbol{\omega}))^2,
\end{equation*}
and
\[
\hat{\mu}_1(\boldsymbol{\omega}) = \frac{\sum_{i=1}^n (1-s_i(\boldsymbol{\omega}))z_i}{\sum_{i=1}^n (1-s_i(\boldsymbol{\omega}))}, \quad
\hat{\mu}_2(\boldsymbol{\omega}) = \frac{\sum_{i=1}^n s_i(\boldsymbol{\omega}) z_i}{\sum_{i=1}^n s_i(\boldsymbol{\omega})}.
\]
Thus, in practice our procedure simply amounts to maximizing Eq.~\eqref{eq:objective_simplified} over $\bm{\omega}$. 

\subsection{Choice of $g$ and optimization tool}
\label{sec:g_choice}

We consider two distinct parameterizations of the boundary function $g(\bm{u};\boldsymbol{\omega})$ and use different optimization tools for each.

\subsubsection{Quadratic polynomial (QP)}
\label{sec:qp}
A natural and interpretable choice for modeling a boundary function is a second-degree polynomial such as 
\begin{equation}
g(\bm{u};\boldsymbol{\omega}) = \bm{u}^{\top} \bm{A} \bm{u} + \bm{b}^{\top} \bm{u} + c, 
\label{eq:qp}
\end{equation} 
with \(\boldsymbol{\omega} = \{\bm{A}, \bm{b}, c\}\). For $\bm{u} \equiv (x,y)^{\top} \in \mathbb{R}^2$, this simplifies to
    \begin{equation*}
        g(x, y;\boldsymbol{\omega}) = ax^2 + bxy + cy^2 + dx + ey + f,
    \end{equation*}
with $\boldsymbol{\omega} = (a, b, c, d, e, f) \in \mathbb{R}^6$ now being a vector of six real-valued coefficients.

This parameterization is expressive enough to represent various boundary shapes and can be interpreted geometrically as a conic section, while still maintaining simplicity. When the goal is to preserve model simplicity, interpretability, and convexity, this choice is particularly suitable.

Since the boundary function \(g(\bm{u}; \boldsymbol{\omega})\) is differentiable with respect to its parameters and lies in a relatively low-dimensional space, we adopt the quasi-Newton method BFGS~\citep[e.g.,][p.~136]{nocedal2006numerical} to optimize the profile log-likelihood. 
    
\subsubsection{Two-layer neural network (2LNN)}
\label{sec:2lnn}
To accommodate more flexible boundary shapes, we can also parameterize the boundary function using a shallow neural network such as 
    \begin{equation}
        g(\bm{u}; \boldsymbol{\omega}) = \boldsymbol{w}_2^\top \tanh(\bm{W}_1 \boldsymbol{u} + \boldsymbol{b}_1) + b_2,
        \label{eq:2lnn}
    \end{equation}
with $\boldsymbol{\omega} = \{\bm{W}_1, \boldsymbol{b}_1, \boldsymbol{w}_2, b_2\}$ being the entire set of trainable parameters (i.e., weights and biases). 
In our current implementation, we use a fully-connected network with one hidden layer of size $32$, i.e., 
$\bm{W}_1 \in \mathbb{R}^{32 \times m}$,
$\bm{b}_1 \in \mathbb{R}^{32}$,
$\bm{w}_2 \in \mathbb{R}^{32}$, and
$b_2 \in \mathbb{R}$. For example, if $\bm{u}\in\mathbb{R}^2$, this leads to an optimization problem in $32 \times 2 + 32 + 32 + 1 = 129$ dimensions. 

This structure introduces nonlinearity through the activation function $\tanh(\cdot)$, enabling the boundary curve to capture complex, non-convex geometries beyond the capacity of a quadratic form. Given the increased number of parameters even for $\bm{u} \in \mathbb{R}^2$ and the added non-convexity of $g$ itself, we employ the ADAM optimizer~\citep{kingma2014adam} for training.

\subsection{Further details}

We now discuss some further practical details. While all our discussions here are applicable to $\bm{u}_i \in \mathbb{R}^m$ for any $m > 1$, to reduce notational clutter such as additional superscripts and subscripts we will restrict $\bm{u}_i \equiv (x_i,y_i)^{\top} \in \mathbb{R}^2$ to be two-dimensional in this section.

\subsubsection{Representative point on the boundary}
\label{sec:representative_point}

While all points on the estimated boundary curve $g(x, y; \hat{\boldsymbol{\omega}}) = 0$ are theoretically equivalent in identifying the beginning of performance plateau, there may be practical situations where selecting a single representative point is desirable.

To determine such a point in a principled manner, we propose an additional heuristic based on the notion of the center of gravity (COG) from classical mechanics.  This approach requires that all performance values $z_i$ have the same sign, which can be ensured through a suitable shift if necessary. Under this assumption, we interpret $|z_i|$ as the ``mass'' located at the point $(x_i, y_i)$ in the parameter space.

\begin{definition}[Center of Gravity (COG)]\label{def:COG}
Let the index sets be defined as
\[
\mathcal{I}_1 := \{i : g(x_i, y_i; \hat{\boldsymbol{\omega}}) \leq 0\}, \quad \mathcal{I}_2 := \{i : g(x_i, y_i; \hat{\boldsymbol{\omega}}) > 0\}.
\]
Then, the center of gravity (COG) in region $\mathcal{I}_j$, for $j \in \{1,2\}$, is the point
$(x_{\text{COG}}^{(j)}, y_{\text{COG}}^{(j)})$, where
\begin{equation}
    x_{\text{COG}}^{(j)} = \frac{\sum_{i \in \mathcal{I}_j} x_i |z_i|}{\sum_{i \in \mathcal{I}_j} |z_i|} \quad\text{and}\quad
    y_{\text{COG}}^{(j)} = \frac{\sum_{i \in \mathcal{I}_j} y_i |z_i|}{\sum_{i \in \mathcal{I}_j} |z_i|}
\end{equation}
are weighted averages of $\{(x_i, y_i): i \in \mathcal{I}_j\}$ using $|z_i|$ as mass.\hfill$\square$
\end{definition}

We then connect the two centers of gravity, $(x_{\text{COG}}^{(1)}, y_{\text{COG}}^{(1)})$ and $(x_{\text{COG}}^{(2)}, y_{\text{COG}}^{(2)})$, with a straight line. The intersection point $(x_b, y_b)$ of this line with the boundary curve $g(x, y; \hat{\boldsymbol{\omega}}) = 0$ is selected as the representative point. 

In practice, it may sometimes be preferable to identify a ``best'' point in the original grid, i.e., one of $\{(x_i,y_i)\}_{i=1}^n$, rather than the interpolated boundary point $(x_b, y_b)$, especially when downstream tasks may require using a model configuration $(x_i, y_i)$ that has already been explicitly evaluated. 
In such cases, users can simply choose the nearest grid point to $(x_b,y_b)$, i.e.,
\begin{equation}
    (x_b^{\dagger}, y_b^{\dagger}) = 
    \underset{i}{\operatorname{arg\,min}} \left\|(x_i, y_i) - (x_b, y_b) \right\|.
\end{equation}
This ensures the selected point is both practically usable and as close as possible to the representative boundary point. 
One can also restrict the choice to the ``good side'' of the boundary, say $\mathcal{I}_j$, and take the ``$\operatorname{arg\,min}$'' above over $i \in \mathcal{I}_j$ alone rather than over all $i$, but such restriction is not necessary. 

\subsubsection{Transformation of optimization landscape}
\label{sec:transform}

In practical applications, the hyper-parameters $(x_i, y_i)$ may exhibit significant skewness or lie on widely different scales. Such disparities in scale can induce a distorted and ill-conditioned geometry in the optimization landscape, potentially hindering the convergence of gradient-based optimizers.

To address these issues, we recommend to apply a convenient yet effective data transformation to each parameter dimension. The transformation consists of a logarithmic scaling---used to compress large values and reduce skewness---followed by standard normalization to produce zero-mean, unit-variance features: 
\begin{equation}
    x_i \leftarrow \frac{\log x_i - \overline{\log x}}{\operatorname{SD}(\log x)}, \qquad
    y_i \leftarrow \frac{\log y_i - \overline{\log y}}{\operatorname{SD}(\log y)},
\end{equation}
where $\overline{\log x}$ and $\operatorname{SD}(\log x)$ denote the sample mean and standard deviation of the log-transformed $x_i$ values (and similarly for $y_i$) over all $i \in \{1,2,\dots,n\}$. 
This preprocessing step helps regularize the parameter space and promotes numerically stable optimization.

\section{Examples}
\label{sec:examples}
We now illustrate the usefulness of our method using real-world examples across a diverse set of modeling contexts, including elastic nets for predicting the size of forest fires, support vector machines for detecting Parkinson's disease, and neural networks for image recognition. 
Although in Section~\ref{sec:main} we have presented our method generally for $m>1$, in reality it is rare for researchers to experimentally decide on more than two hyper-parameters due to the exponential increase in computational cost. The most common practice is to rely on experiments for the two most important---or most sensitive---hyper-parameters, while fixing all others at some pre-determined values.  
Thus in this section, we will consider only two hyper-parameters in all cases as well, i.e., $\bm{u}=(x,y)\in\mathbb{R}^2$; this also makes it easier to examine the results visually. We will display contours of the performance metric, together with the boundary curve $g(x,y;\hat{\bm{\omega}})=0$ computed by our method, the center of gravity on either side of the boundary, and the representative boundary point, etc. Table~\ref{tab:fig_legend} summarizes the legend we use in all of our figures. 

\begin{table}[ht]
\centering
\begin{tabular}{cl}
\hline
 & \\[-12pt]
Legend & \multicolumn{1}{c}{Description (refer to Section~\ref{sec:representative_point})} \\
 & \\[-12pt]
\hline
 & \\[-12pt]
$+$ & center of gravity $(x_{\text{COG}}^{(j)},y_{\text{COG}}^{(j)})$ in $\mathcal{I}_j$, for either $j\in\{1,2\}$\\ 
 & \\[-12pt]
$\times$ & representative boundary point $(x_b, y_b)$\\
 & \\[-12pt]
\tick[rotate=180] & point $(x_b^{\dagger}, y_b^{\dagger})$ in $\mathcal{I}_1$ that is nearest to $(x_b, y_b)$ \\
 & \\[-12pt]
 \tick & point $(x_b^{\dagger}, y_b^{\dagger})$ in $\mathcal{I}_2$ that is nearest to $(x_b, y_b)$ \\
 & \\[-12pt]
\hline
\end{tabular}
\caption{\label{tab:fig_legend}Summary of legend used in Figures~\ref{fig:ff_enet}, \ref{fig:pd_svm}  and \ref{fig:mnist}.}
\end{table}

\subsection{Elastic nets on forest fires data}
\label{sec:example_EN}

Figure~\ref{fig:ENcontour}, which we used in Section~\ref{sec:intro} as a motivation for this study, is the result of fitting the elastic net model \eqref{eq:EN} with different tuning parameters to a dataset from the UCI Machine Learning Repository, listed under the title ``Forest Fires''.  
It contains 517 records collected from the Montesinho natural park in the northeast region of Portugal between January 2000 and December 2003. Each record corresponds to a forest fire occurrence, with 12 input features describing the occurrence (e.g., spatial location and site conditions like temperature, relative humidity, wind speed, etc.) and an outcome variable being the size of the burned area measured in hectares. 

Because the distribution of the outcome variable is highly skewed, as most fires did not cause substantial burns, we apply a logarithmic transformation to it and define our response variable as $y = \ln\{1 + (\text{burned area})\}$. 
Two variables, month and day-of-week, which are nominally coded as categorical in the dataset but clearly have cyclical interpretations in this context, are re-coded as 
$(\cos M, \sin M)$ with $M \in \{(1/12)(2\pi), \dots, (12/12)(2\pi)\}$ for ``jan'', ..., ``dec'' and as 
$(\cos D, \sin D)$ with $D \in \{(1/7)(2\pi), \dots, (7/7)(2\pi)\}$ for ``mon'', ..., ``sun''.

We also make a random 70/30 split of the dataset, using 362 samples for training and the remaining 155 samples for testing. Because different random splits lead to considerable variations in the training and test performances, we do this repeatedly for 10 times, and use only the more stable \emph{average performances} over these 10 repetitions for our discussion.    

\begin{figure}[htp]
    \centering
    \begin{subfigure}[t]{0.48\textwidth}
        \centering
        \includegraphics[width=\linewidth]{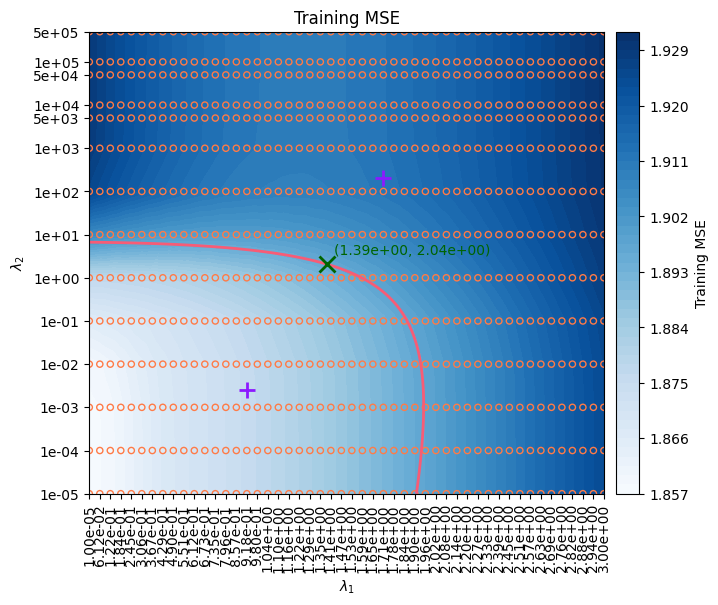}
        \caption{\label{fig:enet_qp_train}QP result + training MSE}
    \end{subfigure}
    \hfill
    \begin{subfigure}[t]{0.48\textwidth}
        \centering
        \includegraphics[width=\linewidth]{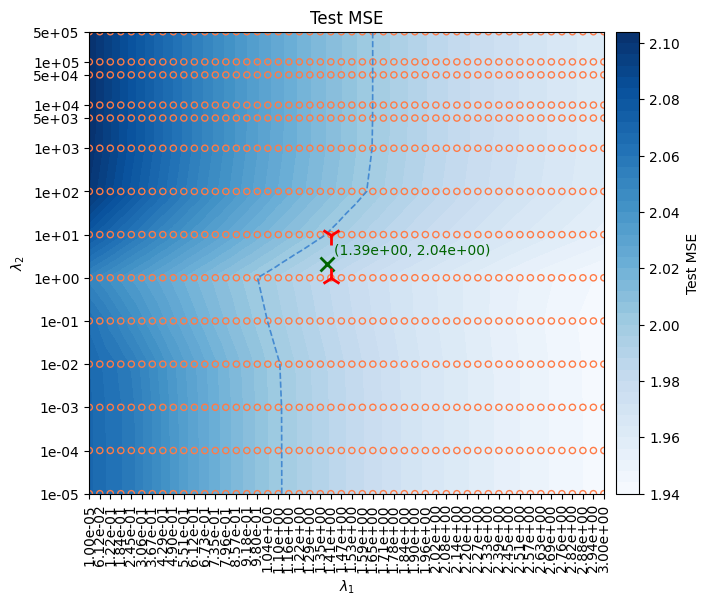}
        \caption{\label{fig:enet_qp_test}QP result + test MSE}
    \end{subfigure} \\[12pt]
    \begin{subfigure}[t]{0.48\textwidth}
        \centering
        \includegraphics[width=\linewidth]{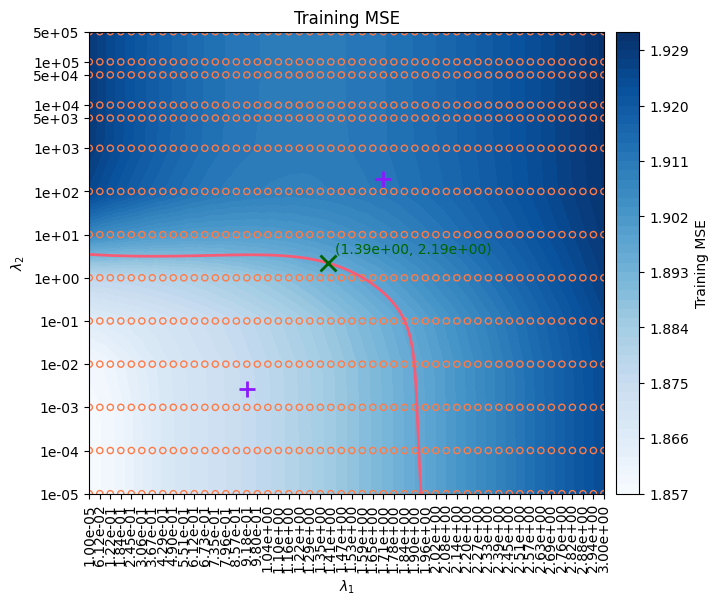}
        \caption{\label{fig:enet_2nn_train}2LNN result + training MSE}
    \end{subfigure}
    \hfill
    \begin{subfigure}[t]{0.48\textwidth}
        \centering
        \includegraphics[width=\linewidth]{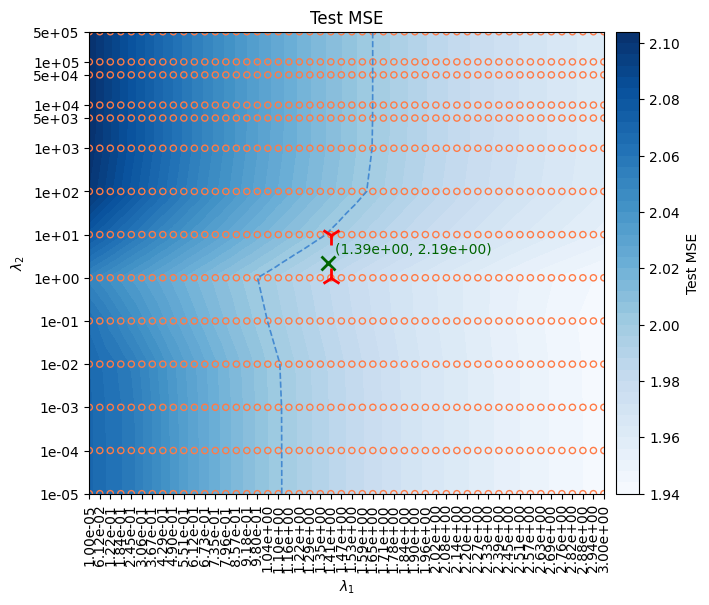}
        \caption{\label{fig:enet_2nn_test}2LNN result + test MSE}
    \end{subfigure}
    \caption{Average mean squared error on the forest fires dataset as a function of two hyper-parameters, $\lambda_1$ and $\lambda_2$.}
    \label{fig:ff_enet}
\end{figure}

We fit elastic nets using the \texttt{enet} function from the R package \texttt{elasticnet}. The left panels of Figure~\ref{fig:ff_enet} show the average training MSE across various $(\lambda_1, \lambda_2)$-pairs. We have already described in Section~\ref{sec:intro} how the training error behaves in this example as the hyperparameters are increased, and these panels now display the boundaries identified by our method, parameterized either as a QP (Figure~\ref{fig:enet_qp_train}) or as a 2LNN (Figure~\ref{fig:enet_2nn_train}). Regardless of which parameterization, the final representative boundary points, labeled ``$\times$'' in all plots, are nearly the same. 

The average test MSE is shown in the right panels of Figure~\ref{fig:ff_enet}. Here, we see that, although the sharp increase in training error, as both $\lambda_1$ and $\lambda_2$ grow beyond a certain point, is a warning against setting either hyperparameter too large, a larger-than-necessary $\lambda_1$ turns out to be relatively harmless in terms of test error. However, the boundary identified by our method essentially nails the critical point at which further increases in $\lambda_2$ will become truly harmful, as well as how far $\lambda_1$ should be allowed to grow for the test error to reach a relatively safe zone. Moreover, the representative boundary point does a good job at simultaneously capturing both of these trade-offs.   

\subsection{Support vector machines on Parkinson data}

Our next example concerns the support vector machine \citep[SVM; see, e.g.,][]{svm95}, a method which dominated the field of machine learning during much of the two decades before the deep learning movement came into full swing. 

Here, we use it to detect the presence of Parkinson’s disease (PD) from acoustic features of voice recordings. 
The dataset originates from \citet{sakar2013parkinson_speech} and is available in the UCI Machine Learning Repository under the title, ``Parkinson's Speech with Multiple Types of Sound Recordings''. It comprises a training set of 1,040 voice samples (sustained vowels, numbers, words, and short sentences), with equal numbers from PD patients and healthy individuals. From each sample, 26 linear and time-frequency based features (e.g., pitch-related measures, number of pulses, degree of voice breaks) are extracted. A separate test set of 168 voice samples that was collected independently under the same clinical protocol provides held-out data for validation.

\begin{table}[htp]
\centering
\fbox{\begin{tabular}{p{0.9\textwidth}}
\begin{enumerate}
\vspace{-\topsep}
\item\label{item:QP} Solve 
\[
\begin{array}{ccl}
\underset{\alpha_1,\dots,\alpha_N}{\max} & & \displaystyle\sum_{i=1}^N \alpha_i - \frac{1}{2} \sum_{i=1}^N\sum_{j=1}^N \alpha_i \alpha_{j} y^{(i)} y^{(j)} K_h(\bm{x}^{(i)};\bm{x}^{(j)}) \\[6mm]
\text{s.t.} & & \displaystyle\sum_{i=1}^N \alpha_i y^{(i)} = 0,\\[6mm]
& & 0 \leq \alpha_i \leq \gamma, \quad \forall\ i \in \{1,2,\dots,N\}.
\end{array}
\]

\item 
Based on the solutions $\alpha_1,\dots,\alpha_N$ to Step~\ref{item:QP},
partition the set $\{1,2,\dots,N\}$ into three disjoint groups:
\[ 
G_0 \equiv \{i: \alpha_i=0\}, \quad
G_{\gamma} \equiv \{i: \alpha_i=\gamma\}, \quad
G_{(0,\gamma)} \equiv \{i: 0 < \alpha_i < \gamma\}.
\]

\item Pick any $i \in G_{(0,\gamma)}$, and solve
\[
\alpha_0 = y^{(i)} - \sum_{j=1}^N \alpha_{j} y^{(j)} K_h(\bm{x}^{(i)};\bm{x}^{(j)}). 
\]
\end{enumerate}
\end{tabular}}
\caption{\label{tab:svmDual}The key steps to fit an SVM for binary classification ($y^{(i)} = \pm 1$), given fixed hyper-parameters $h$ and $\gamma$.}
\end{table}

Given a training dataset $\{(\bm{x}^{(i)}, y^{(i)})\}_{i=1}^N$, where $y^{(i)} = \pm 1$ is a binary label, the SVM classifies a new point $\bm{x}$ by the sign of 
\begin{equation} 
p(\bm{x}) = \sum_{i=1}^N \alpha_{i} y^{(i)} K_h(\bm{x};\bm{x}^{(i)}) + \alpha_0,
\label{eq:svm}
\end{equation} 
where $K_h(\cdot;\cdot)$ is a pre-specified kernel function with hyper-parameter $h$, and $\alpha_0, \alpha_1, \dots, \alpha_N$ are obtained by following the steps given in Table~\ref{tab:svmDual}, which involve an additional hyper-parameter $\gamma$. These steps are nontrivial to explain, but we will spare the readers from such details as they are not directly pertinent to our discussion.

\begin{figure}[htp]
    \centering
    \begin{subfigure}[t]{0.48\textwidth}
        \centering
        \includegraphics[width=\linewidth]{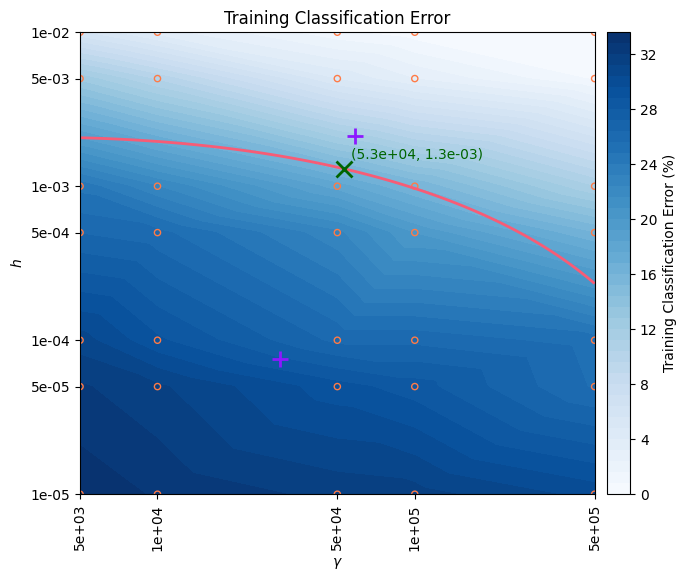}
        \caption{\label{fig:svm_qp_train}QP result + training error}
    \end{subfigure}
    \hfill
    \begin{subfigure}[t]{0.48\textwidth}
        \centering
        \includegraphics[width=\linewidth]{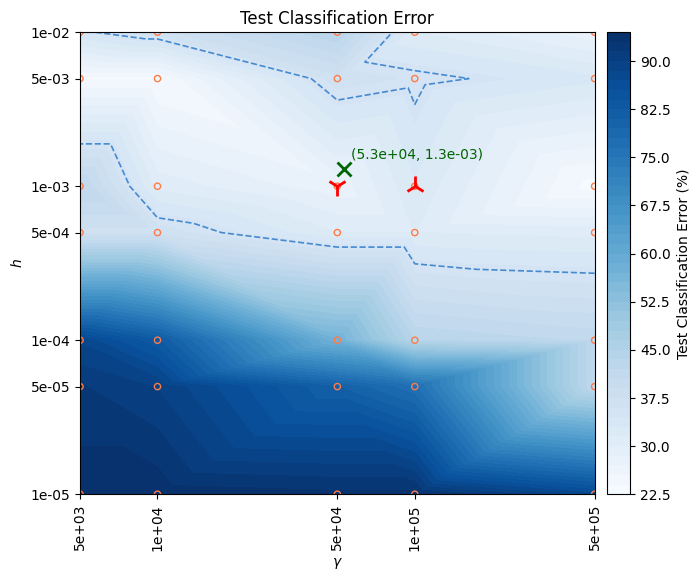}
        \caption{\label{fig:svm_qp_test}QP result + test error}
    \end{subfigure} \\[12pt]
    \begin{subfigure}[t]{0.48\textwidth}
        \centering
        \includegraphics[width=\linewidth]{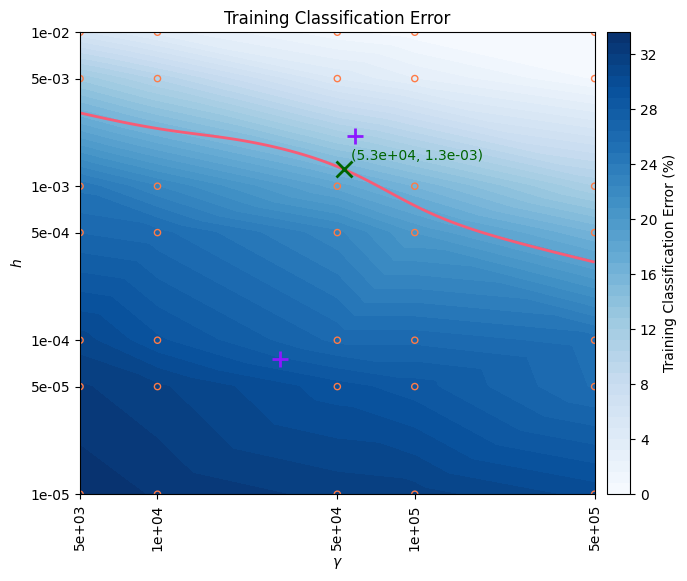}
        \caption{\label{fig:svm_2nn_train}2LNN result + training error}
    \end{subfigure}
    \hfill
    \begin{subfigure}[t]{0.48\textwidth}
        \centering
        \includegraphics[width=\linewidth]{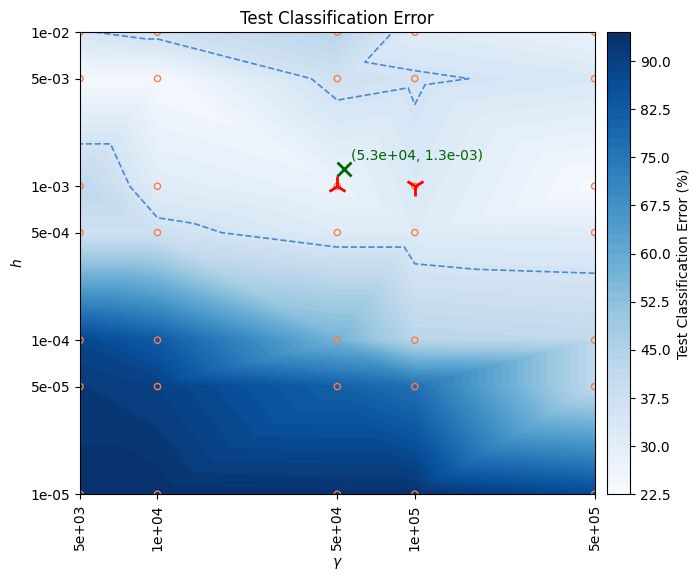}
        \caption{\label{fig:svm_2nn_test}2LNN result + test error}
    \end{subfigure}
    \caption{Classification error (in \%) on the Parkinson dataset as a function of two hyper-parameters, $\gamma$ and $h$; see Table~\ref{tab:svmDual}.}
    \label{fig:pd_svm}
\end{figure}

We use the radial basis kernel,
\begin{equation*}
    K_{h}(\boldsymbol{u}; \boldsymbol{v}) = \exp \left\{-h\|\boldsymbol{u}-\boldsymbol{v}\|^2 \right\},
\end{equation*}
and the \texttt{svm} function from the R package \texttt{e1071} to fit \eqref{eq:svm}.
The left side of Figure~\ref{fig:pd_svm} shows the classification error (in \%) on the training examples as the two hyper-parameters $(\gamma, h)$ vary. The training error starts relatively low in the top-right corner when both $(\gamma, h)$ are relatively large, but these SVMs are overfitting the training data. As their values are decreased, the training error increases, gradually at first but more sharply ``at some point''. 

The left side of the figure also shows the boundaries marking the beginning of these sharp increases as determined by our method, when the boundary itself is parameterized either as a QP (Figure~\ref{fig:svm_qp_train}) or as a 2LNN (Figure~\ref{fig:svm_2nn_train}). Regardless of which parameterization, our method suggests that $\gamma$ should be no less than about $5\times 10^4$ and that $h$ should be no less than about $1\times 10^{-3}$; it also finds almost the same representative boundary point at roughly $(\gamma_b, h_b)=(5.3\times 10^4, 1.3\times 10^{-3})$, labeled ``$\times$'' in all panels of Figure~\ref{fig:pd_svm}.
  
The right side of Figure~\ref{fig:pd_svm} shows the classification error (again, in \%) on the 168 held-out samples. As expected, the error surface is noisier on held-out data, but it is reassuring to see that the representative boundary point at roughly $(\gamma_b, h_b)=(5.3\times 10^4, 1.3\times 10^{-3})$ clearly corresponds to an SVM which would have a favorably low test error. Although such an SVM has not been fitted, among those already fitted the ``closest'' one is at $(\gamma_b^{\dagger}, h_b^{\dagger})=(5\times 10^4,1\times 10^{-3})$. Its test error of about $27.38\%$ is indeed among the lowest of all SVMs fitted. In Figures~\ref{fig:svm_qp_test} and \ref{fig:svm_2nn_test}, it is labeled ``\tick[rotate=180]'' and ``\tick'', respectively. 

\subsection{Neural networks on MNIST data}
\label{sec:example_mnist}

Our final example is an application of our method to selecting hyper-parameters when training neural networks, based on the paper by \citet{srivastava14}. In the rapidly evolving field of deep learning, this paper is considered an early contribution but it remains an influential classic for popularizing the so-called ``dropout'' technique, whose combination with another procedure called ``max-norm regularization'' was found to be especially effective at reducing generalization error \citep[Section 6.5]{srivastava14}. 

To illustrate, we use the well-known MNIST dataset, which contains 60,000 training examples and 10,000 test examples, each representing a $28 \times 28$ grayscale image of a handwritten digit. The classification task is to assign each image to one of the ten digit classes, \{0, 1, \dots, 9\}.

\begin{table}[htp]
\centering
\fbox{\begin{tabular}{p{0.9\textwidth}}
Initialize $\bm{\theta} := \{\bm{W}_1, \bm{W}_2, \bm{W}_3, \bm{b}_1, \bm{b}_2, \bm{b}_3\}$ (more details in Appendix~\ref{app:mnist_config}) and set $\bm{m}_0 \leftarrow \bm{0}$. \\[1mm] 
For $t = 1, \ldots, T$:
\begin{enumerate}
    \item\label{step:samle} Sample a mini-batch $\{(\bm{x}^{(i)}, y^{(i)})\}^B_{i=1}$ from the training dataset $\mathcal{D}$.
    \item\label{step:grad} Calculate  
    $\bm{g}_t \leftarrow \nabla_{\bm{\theta}} \tilde{L}_t (\bm{\theta}_{t-1})$, where 
    \[
    \tilde{L}_t (\bm{\theta}) = 
    \frac{1}{B} \sum_{i=1}^B L\left(y^{(i)}, \mathbb{D}_t^{(i)}[\bm{p}(\bm{x}^{(i)}; \bm{\theta}); r]\right)   
    \]
    and $\mathbb{D}_t^{(i)}(\cdot; r)$ is a stochastic operation which, for every $i$ and $t$, randomly ``drops out'' a set of nodes in its neural network argument and rescales the remaining ones (more details in Appendix~\ref{app:mnist_dropout}). (This makes $\bm{g}_t$ here slightly different from $\nabla_{\bm{\theta}} L_t (\bm{\theta}_{t-1})$, where 
    \[
    L_t(\bm{\theta}) = 
    \frac{1}{B} \sum_{i=1}^B L\left(y^{(i)}, \bm{p}(\bm{x}^{(i)}; \bm{\theta})\right)
    \]
    is the ``usual'' average loss over the mini-batch.)
    \item\label{step:update} Update 
    \[
    \bm{m}_t \leftarrow \mu_t \bm{m}_{t-1} + \bm{g}_t
    \quad\text{and}\quad
    \bm{\theta}_t \leftarrow \bm{\theta}_{t-1} - \epsilon_t \bm{m}_t
    \]
    where $\mu_t$ and $\epsilon_t$ each follow a pre-specified schedule over $t$ (more details in Appendix~\ref{app:mnist_config}).
    \item\label{step:max_norm} Rescale each row, $\bm{w}_j$, of $\bm{W}_1, \bm{W}_2, \bm{W}_3 \in \bm{\theta}_t$ by
    \[
    \bm{w}_j \leftarrow \bm{w}_j \cdot \min\left(1, \frac{c}{\|\bm{w}_j\|_2 + \varepsilon}\right), \quad \varepsilon = 10^{-7}.
    \]
\end{enumerate}
End For
\end{tabular}}
\caption{\label{tab:SGD}Training the neural network in Eq.~\eqref{eq:mnist_net} using a stochastic gradient descent algorithm with both dropout (the $\mathbb{D}_t^{(i)}$ operator with dropout rate $r$ in Step~\ref{step:grad}) and max-norm regularization (the $c$ parameter in Step~\ref{step:max_norm}).}
\end{table}

We use a stochastic gradient descent algorithm (Table~\ref{tab:SGD}) with both dropout (Step~\ref{step:grad} of Table~\ref{tab:SGD}) and max-norm regularization (Step~\ref{step:max_norm} of Table~\ref{tab:SGD}) to train a neural network $\bm{p}(\bm{x})$ by the cross-entropy loss function, 
\[
L(y, \bm{p}(\bm{x})) =  - \log(p_{y}(\bm{x})), \quad  y \in \{1,2,\dots,K\},  
\]
where $p_k(\bm{x})$ denotes the $k$-th element of $\bm{p}(\bm{x}) \in (0,1)^K$. 
Statisticians may simply know this loss function better as the negative multinomial log-likelihood. 
For $\bm{p}(\bm{x})$, we use one of the architectures considered by \citet[][Section 6.1]{srivastava14} with two hidden layers (more details in Appendix~\ref{app:mnist}), 
\begin{equation}
\bm{p}(\bm{x}) = \sigma \left[\bm{W}_3 
  \rho \left\{ \bm{W}_2 
   \rho( \bm{W}_1 \bm{x}  + \bm{b}_1 ) + \bm{b}_2
       \right\} + \bm{b}_3
                 \right], 
\label{eq:mnist_net}
\end{equation}
where $\rho(\bm{z})=\max(\bm{0},\bm{z})$ is the ReLU activation function applied element-wise, and $$\sigma(z_1,\dots,z_K)=\frac{1}{(e^{z_1}+\dots+e^{z_K})}(e^{z_1},\dots,e^{z_K})$$ is the softmax function that maps $\mathbb{R}^K$ to $(0,1)^K$. For the MNIST data, $K=10$ and $\bm{x} \in \mathbb{R}^{784}$.  

\begin{figure}[htp]
    \centering
    \begin{subfigure}[t]{0.48\textwidth}
        \includegraphics[width=\linewidth]{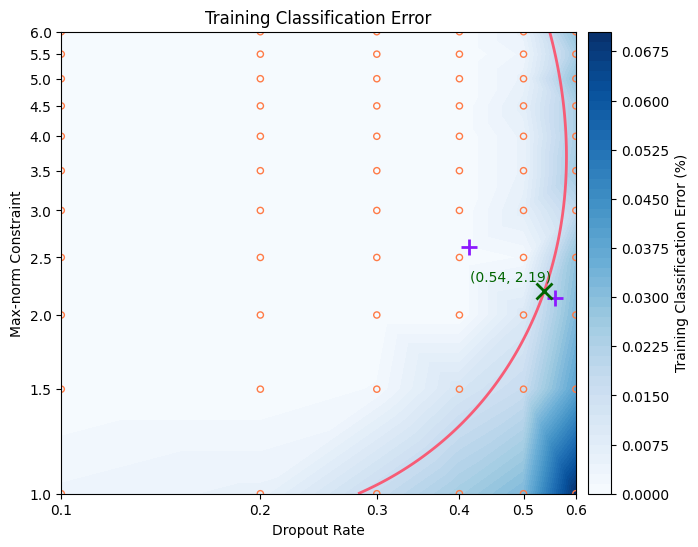}
        \caption{\label{fig:mnist_qp_train}QP result + training error}
    \end{subfigure}
    \hfill
    \begin{subfigure}[t]{0.48\textwidth}
        \includegraphics[width=\linewidth]{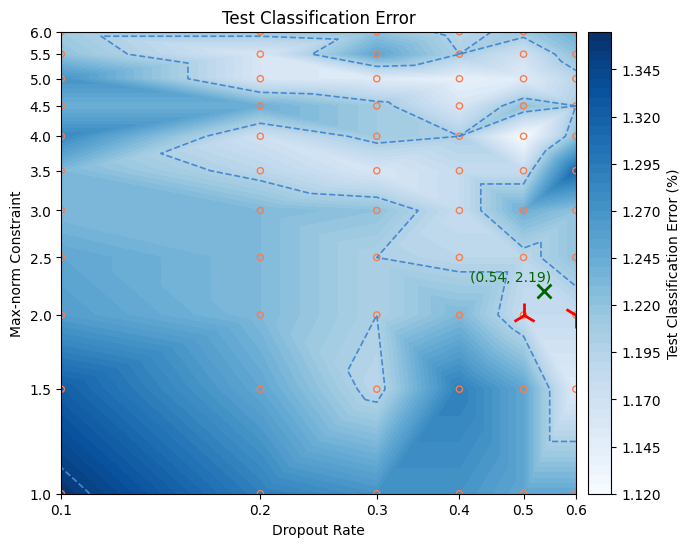}
        \caption{\label{fig:mnist_qp_test}QP result + test error}
    \end{subfigure}\\[12pt]
    \begin{subfigure}[t]{0.48\textwidth}
        \includegraphics[width=\linewidth]{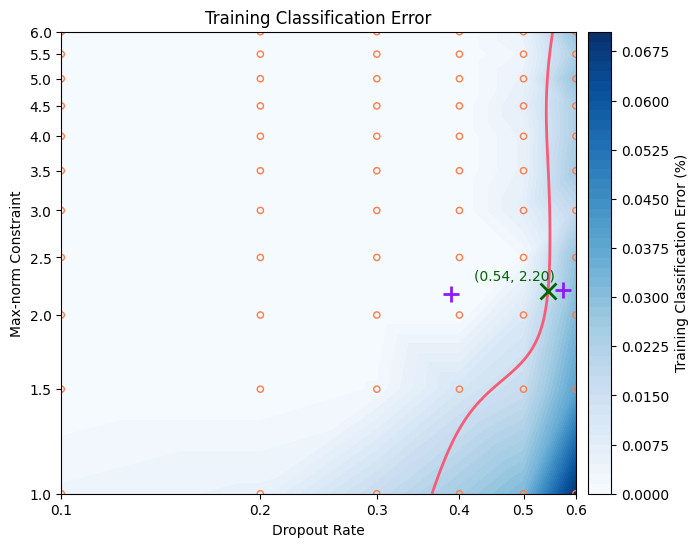}
        \caption{\label{fig:mnist_2nn_train}2LNN result + training error}
    \end{subfigure}
    \hfill
    \begin{subfigure}[t]{0.48\textwidth}
        \includegraphics[width=\linewidth]{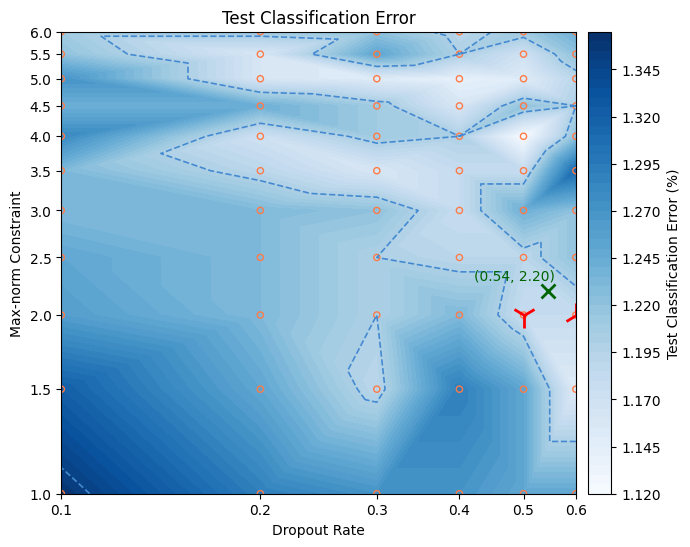}
        \caption{\label{fig:mnist_2nn_test}2LNN result + test error}
    \end{subfigure}
    \caption{Classification error (in \%) on the MNIST dataset as a function of the dropout rate $r$ and the max-norm parameter $c$; see Table~\ref{tab:SGD}.}
    \label{fig:mnist}
\end{figure}

The left side of Figure~\ref{fig:mnist} shows the classification error (in \%) on the training examples as the dropout parameter $r$ (Step~\ref{step:grad} of Table~\ref{tab:SGD}) and the max-norm parameter $c$ (Step~\ref{step:max_norm} of Table~\ref{tab:SGD}) vary over a ``critical range''. The training error starts relatively low in the top-left corner when $r$ is small and $c$ is large, but the resulting networks in that region are likely overfitting the training data. As we gradually increase $r$ and decrease $c$, the training error increases, slowly at first but more abruptly ``at some point''. 

Also shown on the left side of the figure are the boundaries marking the beginning of these abrupt increases as determined by our method, when the boundary itself is parameterized either as a QP (Figure~\ref{fig:mnist_qp_train}) or as a 2LNN (Figure~\ref{fig:mnist_2nn_train}). Using either parameterization, our method suggests a dropout rate of no more than about $r = 0.55$ and a max-norm parameter of no less than about $c=2.0$; it also finds an almost identical representative boundary point at roughly $(r_b,c_b)=(0.54,2.2)$, labeled ``$\times$'' in Figure~\ref{fig:mnist} throughout.
  
The right side of Figure~\ref{fig:mnist} shows the classification error (again, in \%) on the 10,000 test examples. As is always the case, the overall performance landscape is much noisier on the test data. The reaffirming observation here is that the representative boundary point at roughly $(r_b,c_b)=(0.54,2.2)$ does indeed lie in a region of relatively low test error. Although a neural network has not been trained with this particular combination (i.e., a dropout rate of $0.54$ and a max-norm parameter of $2.2$), among those already trained the ``nearest'' one is at $(r_b^{\dagger}, c_b^{\dagger})=(0.5,2.0)$---labeled ``\tick'' in Figure~\ref{fig:mnist_qp_test} and ``\tick[rotate=180]'' in Figure~\ref{fig:mnist_2nn_test}; it has a competitive low test error of about $1.18\%$. 

\section{Discussion}
\label{sec:disc}

We now make a few general observations and remarks.  

First, the difference between the two parameterizations of the boundary (Section~\ref{sec:g_choice}) is well within expectation: the QP parameterization (Section~\ref{sec:qp}) gives rise to boundaries that are highly regular, convex, and quadratic, while the 2LNN parameterization (Section~\ref{sec:2lnn}) leads to boundaries that are more flexible, but also more irregular, and not necessarily convex. 
However, despite these differences in appearance, in terms of dividing the points $\{\bm{u}_1, \bm{u}_2, \dots, \bm{u}_n\}$ into two regions the two types of boundaries are still in high agreement with each other. 
The level of agreement between the two parameterizations is even higher when it comes to the representative boundary point (Section~\ref{sec:representative_point}). This is also meant as the default answer of our method; in other words, it is our method's automatic choice of the hyper-parameters. As such, it is worth noting that this choice also often turns out to be a practically good one.  

Next, the choice of $f(z;\bm{\theta}_j)$ simply as Gaussian with a common variance for $j \in \{1,2\}$ is inherited from the parent work \citep{zhu2006} of this project, and we have noted earlier in Section~\ref{sec:f_choice} that it is quite arbitrary. 
But any method will naturally work well when its assumption is satisfied, however arbitrary the assumption may be. In our case, this will happen if Eq.~\eqref{eq:f_choice} is a reasonably good model for $z_1, z_2, \dots, z_n$. We have found this to be the case usually, if one focuses correctly on the critical region of the hyper-parameters where the model performance is changing the most, but poorly focused searches (of hyper-parameters) can certainly cause $z_1, z_2, \dots, z_n$ to deviate severely from such an assumption and reduce the method's effectiveness as a result. However, we are able to confirm with a simulation study (see Appendix~\ref{app:sim}) that, as long as $z_1, z_2, \dots, z_n$ can be reasonably described by two distributions $f(z;\bm{\theta}_1)$ and $f(z;\bm{\theta}_2)$ with a distinct locational shift, modeling them as Gaussian, even if the true distributions are not, does not lead to substantial performance deterioration of our method.

Finally, readers may have noticed that, in Section~\ref{sec:examples}, our $z_1, z_2, \dots, z_n$ were often simply training performances at different hyper-parameter configurations, rather than properly cross-validated performances. While this clearly saves the amount of computation needed, our method is also particularly conducive to such practical shortcuts because, instead of aiming for configurations which optimize model performance, our method aims for configurations ``at which improvements in model performance can be considered to have plateaued''. In the former context, only model performances measured independently of the training data are meaningful; whereas, in the latter, model performances measured on the training data itself can be meaningful, too.

\section{Other applications}
\label{sec:other_appl}

We end by noting that, in addition to hyper-parameter selection, we may also encounter other practical instances where ``a boundary in the space of [explanatory variable $\bm{u}$] at which [changes in a certain target variable $z$] can be considered to have plateaued'' is of direct interest. For example, surely we would all be happier with more wealth and freedom, but it is also conceivable that their marginal effects on our happiness will likely diminish beyond a certain point, and we may be interested in where such a boundary lies. 

As an illustration of this phenomenon, we single out three national-level variables from the most recent World Happiness Report~\citep[][Figure~2.1]{happy25}: for happiness, we use the three-year average national response to a Gallup World Poll (GWP) question about life evaluation; for wealth, we use log-GDP per capita; and for freedom, we use the average national response to a GWP question about freedom to make life choices. The latter two variables---about wealth and freedom---have both been rescaled by the authors of the report; for details about the nature of their rescaling, refer to various parts of \citet{happy25}, e.g., their Box 2.1, Table 2.1, Box 2.2, and Chapter 2 Statistical Appendix. These three variables from a total of 144 countries (see Appendix~\ref{app:other_appl}) are plotted in Figures~\ref{fig:happiness_qp} and \ref{fig:happiness_nn}, together with the boundaries identified by our method. The boundaries clearly indicate that, while a country's GDP must reach a certain threshold for its citizens' life evaluation to be among the leading group, a wider perception of freedom to choose can, in fact, lower this threshold. 

A more classic example is provided by \citet[][pp.~395--396]{bh2text}, who described a small experimental study of harmful emissions from automobile engines. The amount of carbon monoxide (CO) emission was measured against two design variables: added ethanol and air-to-fuel ratio (AFR), each at three levels---low ($-1$), medium ($0$), and high ($+1$). The data here generally point to a very strong interaction between the two: CO emission is highest when \emph{both} the ethanol level is high \emph{and} the AFR is low, and lowest when \emph{either} the ethanol level is low \emph{or} the AFR is high. This was confirmed by their Figure 11.17 \citep[][p.~466]{bh2text}, and is also evident from our Figures~\ref{fig:co_qp} and \ref{fig:co_nn}, where the boundaries identified by our method, at which drops in CO emission can be considered to have stabilized, are also shown. 

\begin{figure}[htp]
    \centering
    \begin{subfigure}[t]{0.48\textwidth}
        \includegraphics[width=\linewidth]{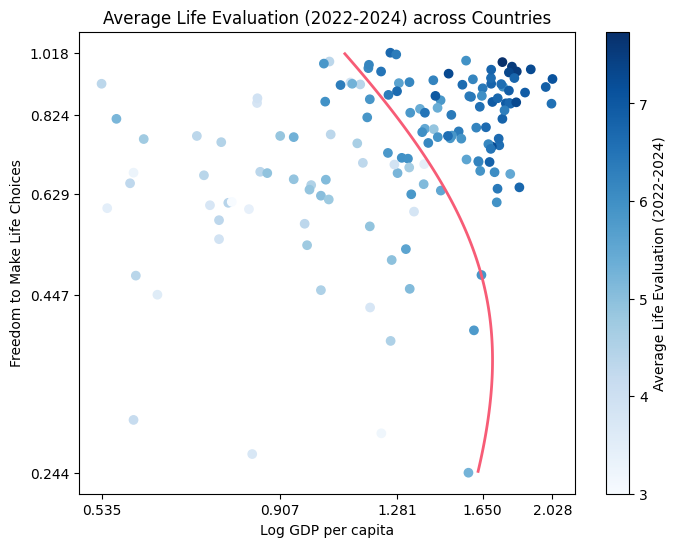}
        \caption{\label{fig:happiness_qp}QP result, Happiness Data}
    \end{subfigure}
    \hfill
    \begin{subfigure}[t]{0.48\textwidth}
        \includegraphics[width=\linewidth]{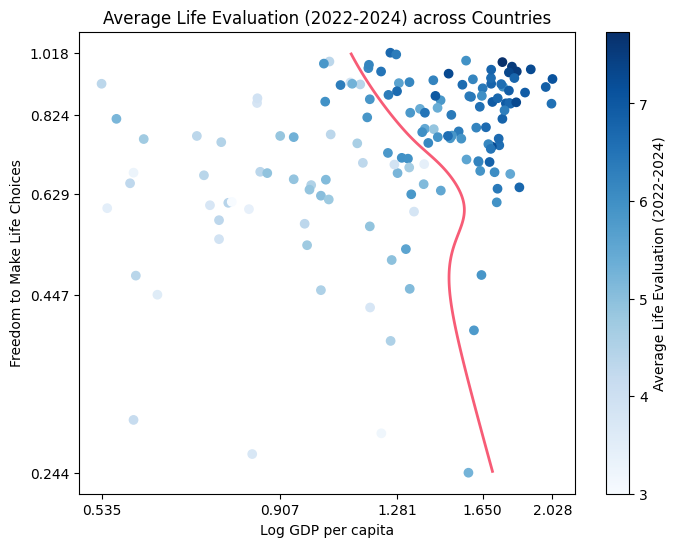}
        \caption{\label{fig:happiness_nn}2LNN result, Happiness Data}
    \end{subfigure}\\[12pt]
    \begin{subfigure}[t]{0.48\textwidth}
        \includegraphics[width=\linewidth]{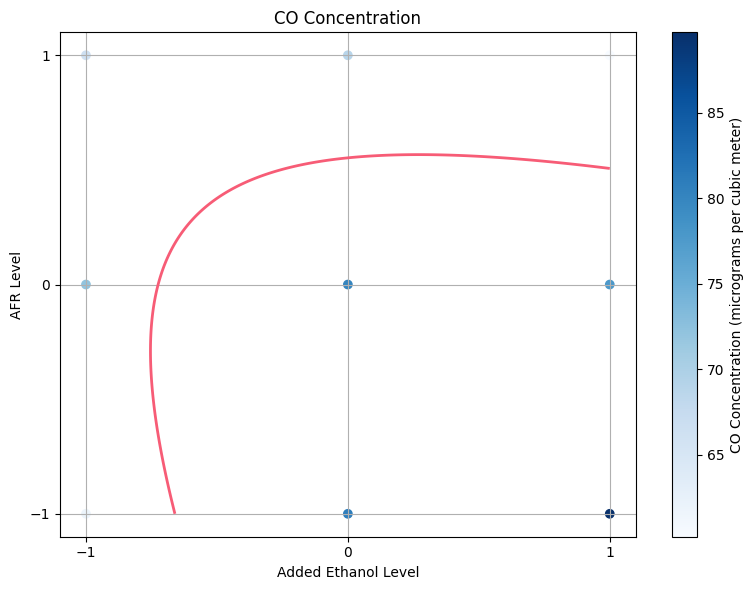}
        \caption{\label{fig:co_qp}QP result, CO Emission Data}
    \end{subfigure}
    \hfill
    \begin{subfigure}[t]{0.48\textwidth}
        \includegraphics[width=\linewidth]{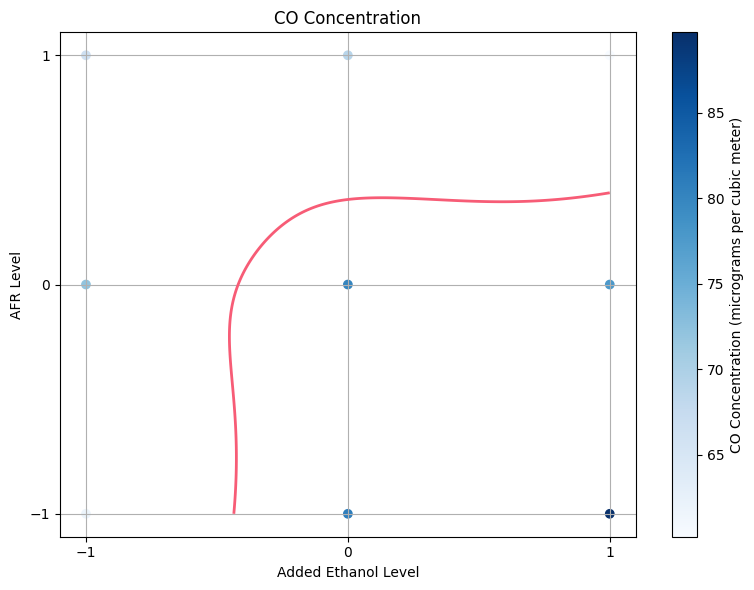}
        \caption{\label{fig:co_nn}2LNN result, CO Emission Data}
    \end{subfigure}\\[12pt]
     \caption{\label{fig:other_appl}Applications of our method to other data-analytic tasks---top, happiness as a function of wealth and freedom; bottom, carbon monoxide (CO) emission as a function of added ethanol and air-to-fuel ratio (AFR).}
\end{figure}

A few remarks we made in Section~\ref{sec:disc} are worth reiterating. First, the boundaries have different appearances depending on how they are parameterized (QP versus 2LNN), but they agree in terms of how the space should be divided---here, into two regions of high versus low life evaluations, and of high versus low CO emission. Next, the main reason why our method works well here is, again, driven mostly by the fact that Eq.~\eqref{eq:f_choice} turns out to be a reasonably good model for these data (see Figure~\ref{fig:Hist} in Appendix~\ref{app:other_appl}). While we believe this can hold in some other practical applications too, we certainly do not anticipate it to hold universally for all potential applications. However, the simulation study reported in Appendix~\ref{app:sim} confirms a certain level of robustness to this model; and in principle one is also free to make a different model assumption in Eq.~\eqref{eq:f_choice} and still apply our methodology to find the desired boundary by maximizing the softened profile log-likelihood~\eqref{eq:objective_soft}. 

\appendix

\section{Derivation of Eq.~\eqref{eq:objective_simplified}}
\label{app:A}

In this appendix we show that, under assumption \eqref{eq:f_choice}, Eq.~\eqref{eq:objective_soft} can be further reduced to Eq.~\eqref{eq:objective_simplified}. First, it is elementary to establish that, under assumption \eqref{eq:f_choice} with $\bm{\theta}_j \equiv (\mu_j, \sigma^2)$ for $j \in \{1,2\}$, the solutions to \eqref{eq:thetaMLE} while fixing $\bm{\omega}$ are given by 
\[
\hat{\mu}_1(\boldsymbol{\omega}) = \frac{\sum_{i=1}^n (1-s_i(\boldsymbol{\omega}))z_i}{\sum_{i=1}^n (1-s_i(\boldsymbol{\omega}))}, \quad
\hat{\mu}_2(\boldsymbol{\omega}) = \frac{\sum_{i=1}^n s_i(\boldsymbol{\omega}) z_i}{\sum_{i=1}^n s_i(\boldsymbol{\omega})}, \quad
\hat{\sigma}^2(\boldsymbol{\omega}) = \frac{\mathrm{RSS}_1(\boldsymbol{\omega}) + \mathrm{RSS}_2(\boldsymbol{\omega})}{n},
\]
where 
\[
\mathrm{RSS}_1(\boldsymbol{\omega}) = 
\sum_{i=1}^n (1-s_i(\boldsymbol{\omega})) (z_i - \hat{\mu}_1(\boldsymbol{\omega}))^2, 
\quad 
\mathrm{RSS}_2(\boldsymbol{\omega}) = 
\sum_{i=1}^n s_i(\boldsymbol{\omega}) (z_i - \hat{\mu}_2(\boldsymbol{\omega}))^2. 
\]
Plugging these back into Eq.~\eqref{eq:objective_soft}, we now get
\begin{align*}
    \ell^{\text{soft}}_{\bm{\omega}}(\bm{\omega}) &= \sum_{i=1}^n (1-s_i(\bm{\omega}))\log f\big(z_i;\hat{\boldsymbol{\theta}}^{\text{soft}}_1(\boldsymbol{\omega})\big) + \sum_{i=1}^n s_i(\boldsymbol{\omega}) \log f\big(z_i;\hat{\boldsymbol{\theta}}^{\text{soft}}_2(\boldsymbol{\omega})\big)\\
    &= \sum_{i=1}^n (1-s_i(\bm{\omega})) \log \left\{ \frac{1}{\sqrt{2\pi \hat{\sigma}^2(\boldsymbol{\omega})}} \exp\left\{-\frac{(z_i - \hat{\mu}_1(\boldsymbol{\omega}))^2}{2\hat{\sigma}^2(\boldsymbol{\omega})}\right\}\right\} + \\
    & \hspace{1.75in} 
    \sum_{i=1}^n s_i(\bm{\omega}) \log \left\{ \frac{1}{\sqrt{2\pi \hat{\sigma}^2(\boldsymbol{\omega})}} \exp\left\{-\frac{(z_i - \hat{\mu}_2(\boldsymbol{\omega}))^2}{2\hat{\sigma}^2(\boldsymbol{\omega})}\right\}\right\}\\
    &= \sum_{i=1}^n (1-s_i(\bm{\omega})) \left\{ -\frac{1}{2}\log(2\pi\hat{\sigma}^2(\bm{\omega})) -\frac{(z_i - \hat{\mu}_1(\boldsymbol{\omega}))^2}{2\hat{\sigma}^2(\boldsymbol{\omega})}\right\} + \\ 
    & \hspace{1.75in} 
    \sum_{i=1}^n s_i(\bm{\omega}) \left\{ -\frac{1}{2}\log(2\pi\hat{\sigma}^2(\bm{\omega})) -\frac{(z_i - \hat{\mu}_2(\boldsymbol{\omega}))^2}{2\hat{\sigma}^2(\boldsymbol{\omega})}\right\}\\
    &= -\frac{1}{2}n\log(2\pi\hat{\sigma}^2(\bm{\omega})) - \\
    & \hspace{1in}
    \frac{1}{2\hat{\sigma}^2(\boldsymbol{\omega})}\left\{\sum_{i=1}^n (1-s_i(\bm{\omega}))(z_i - \hat{\mu}_1(\boldsymbol{\omega}))^2 + \sum_{i=1}^n s_i(\bm{\omega})(z_i - \hat{\mu}_2(\boldsymbol{\omega}))^2 \right\}\\
    &= -\frac{1}{2}n\log(2\pi\hat{\sigma}^2(\bm{\omega})) -\frac{1}{2\hat{\sigma}^2(\boldsymbol{\omega})}\left\{\mathrm{RSS}_1(\boldsymbol{\omega}) + \mathrm{RSS}_2(\boldsymbol{\omega})\right\}\\
    &= -\frac{n}{2}\log(2\pi\hat{\sigma}^2(\bm{\omega})) -\frac{n}{2},
\end{align*}
which is Eq.~\eqref{eq:objective_simplified}.

\section{Details of the MNIST example in Section~\ref{sec:example_mnist}}
\label{app:mnist}

For the MNIST data, each input is a $28\times28$ image represented as $\bm{x} \in \mathbb{R}^{784}$ ($28\times28=784$). The neural network architecture used in Section~\ref{sec:example_mnist} is one considered by
\citet{srivastava14} with two hidden layers, each consisting of 2048 nodes; thus, 
$$ \begin{array}{ll}
\bm{W}_1 \in \mathbb{R}^{2048 \times 784},  & \bm{b}_1 \in \mathbb{R}^{2048}, \\
\bm{W}_2 \in \mathbb{R}^{2048 \times 2048}, & \bm{b}_2 \in \mathbb{R}^{2048}, \\
\bm{W}_3 \in \mathbb{R}^{10 \times 2048},   & \bm{b}_3 \in \mathbb{R}^{10}. 
\end{array}
$$

\subsection{Details about dropout}
\label{app:mnist_dropout}

Following \citet{srivastava14}, we only study the impact of the dropout rate $r$ used for the hidden layers, while for the input layer the dropout rate is fixed at $0.2$. The exact specification of our dropout operator $\mathbb{D}(\cdot; r)$, when applied to Eq.~\eqref{eq:mnist_net}, is:
\[
\mathbb{D}(\bm{p}(\bm{x}); r) = \sigma\left[\bm{W}_3 D_{r}
 \left\{\rho \left(\bm{W}_2 \left[D_{r} 
 \left\{\rho \left(\bm{W}_1 \left[D_{0.2}(\bm{x}) \right] + \bm{b}_1 \right) 
 \right\} \right] + \bm{b}_2 \right) \right\} + \bm{b}_3\right],
\]
where
$$D_{r}(\bm{z}) = \frac{\bm{1}_r \odot \bm{z}}{1-r};$$ 
$\bm{1}_r$ is an i.i.d.~vector drawn element-wise from the Bernoulli$(1-r)$ distribution; and ``$\odot$'' denotes element-wise multiplication. That is, $D_r(\bm{z})$ randomly annihilates some elements of $\bm{z}$ by turning them into zeros, and rescales the rest by one minus the dropout rate, so that its output is equal \emph{in expectation} to its input:  
$$\mathbb{E}(D_r(\bm{z})) = \frac{\mathbb{E}(\bm{1}_r) \odot \bm{z}}{1-r} = \bm{z}.$$

\subsection{Other details}
\label{app:mnist_config}

For the MNIST data, the stochastic gradient descent algorithm in Table~\ref{tab:SGD} initializes
the vectors $\bm{b}_1, \bm{b}_2, \bm{b}_3$ to $\bm{0}$, and draws the initial weight matrices
\[
\bm{W}_1 \sim \mathcal{N}\left(0, \frac{1}{\sqrt{784}}\right), \quad
\bm{W}_2, \bm{W}_3 \sim \mathcal{N}\left(0, \frac{1}{\sqrt{2048}}\right) 
\] 
element-wise from scaled normals, all centered at zero. Furthermore, it uses
\begin{itemize}
\item a total of $T=1,000,000$ iterations;
\item a mini-batch size of $B=100$;
\item a momentum schedule of
    \[
    \mu_t = 
    \begin{cases}
        0.9 + 0.05 \cdot (t-1)/10,000, & t \leq 10,000, \\
        0.95, & t > 10,000;
    \end{cases}
    \]
\item and a learning-rate schedule of
    \[
    \epsilon_t = \epsilon_0 \cdot \gamma^{t-1} 
    \quad\text{with}\quad \epsilon_0 = 0.01 
    \quad\text{and}\quad \gamma = 0.5^{1/10,000}.
    \]
\end{itemize}

\section{Details of other applications in Section~\ref{sec:other_appl}}
\label{app:other_appl}

For the year of 2024, the data set from \citet[][Figure 2.1]{happy25} contains life evaluations for 147 countries. Three countries---Afghanistan, Tajikistan and Venezuela---have been omitted from Figures~\ref{fig:happiness_qp} and \ref{fig:happiness_nn}, because their ``GDP variable'' or ``freedom variable'' is either missing or has been rescaled by the authors to $0.000$, which prevents us from applying the logarithmic transform prior to optimization (see Section~\ref{sec:transform}). 

\begin{figure}[htp]
    \centering
    \begin{subfigure}[t]{0.48\textwidth}
    \includegraphics[width=\linewidth]{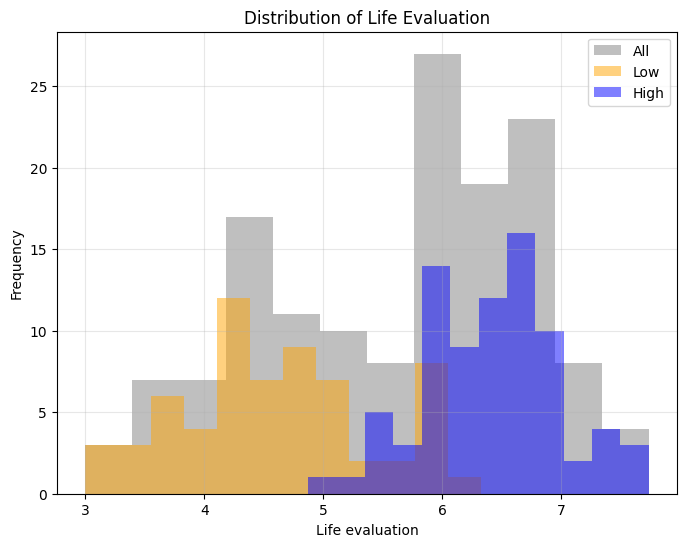}
    \caption{\label{fig:happyHist}Happiness Data}
    \end{subfigure}
    \hfill
    \begin{subfigure}[t]{0.48\textwidth}
    \includegraphics[width=\linewidth]{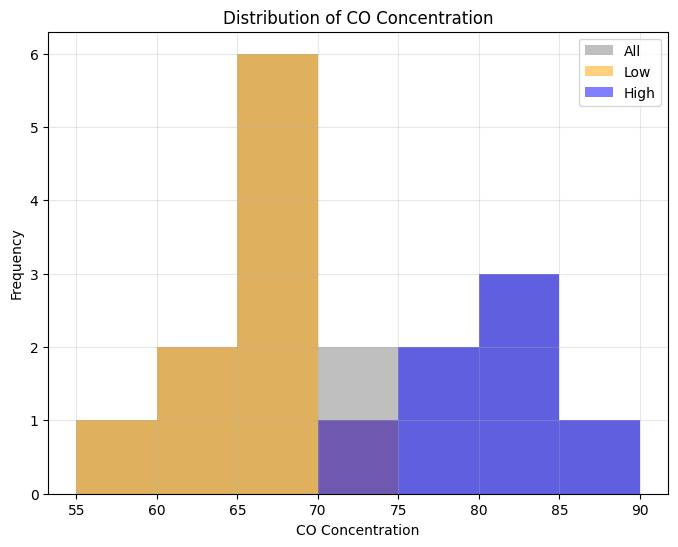}
    \caption{\label{fig:coHist}CO Emission Data}
    \end{subfigure}    
    \caption{\label{fig:Hist}Distribution of the target variable---life evaluation (\ref{fig:happyHist}) and CO concentration (\ref{fig:coHist})---on either side of the boundary (Figure~\ref{fig:other_appl}).}
\end{figure}

Figure~\ref{fig:Hist} shows that Eq.~\eqref{eq:f_choice} is not a grossly inappropriate model for the ``happiness'' data despite a noticeable second mode in the distribution on the left, and that it is even a pretty good one for the ``CO emission'' data. 

\section{Violations to the normality assumption}
\label{app:sim}

In this appendix, we conduct a small simulation study and investigate violations to the normality assumption made by Eq.~\eqref{eq:f_choice}. First, we generate $(x_1,y_1), \dots, (x_{100},y_{100})$ uniformly on $(0,2\pi) \times (-5,5)$. Then, for each $i \in \{1, \dots, 100\}$, we draw 
\[
z_i \sim \begin{cases} 
f(z;\bm{\theta}_1), & \text{if} \quad g(x_i,y_i) \equiv y_i - \sin(2x_i) \leq 0; \\
f(z;\bm{\theta}_2), & \text{otherwise}. 
\end{cases}
\] 
This is not exactly the ``right'' generating mechanism for the kind of data to which our method is usually applied. For example, in our applications the value of $z_i$ will typically fluctuate much less haphazardly if sorted by how far the point $(x_i,y_i)$ is from the boundary $g(x,y)=0$, but this simpler generating mechanism will suffice for our investigation here.

We create a violation of assumption \eqref{eq:f_choice} by taking $f(z;\bm{\theta}_j)$ to be a much heavier-tailed, Laplace density function,
\begin{equation}
f(z; \bm{\theta}_j) = \frac{1}{2\xi} \exp\left\{-\frac{|z-\mu_j|}{\xi}\right\}, \quad j \in \{1,2\},
\label{eq:f_sim}
\end{equation}
with $\xi=1$, $\mu_1=0$ and varying $\mu_2$ over $\{3,2,1\}$. We then assess how well our algorithm can estimate $g(x,y)$ by the fraction of points that are incorrectly assigned by our estimated boundary to the wrong side. (The smaller this fraction is, the better.) 

The same metric is then compared against a benchmark scenario which draws $z_i$ from Eq.~\eqref{eq:f_choice} rather than Eq.~\eqref{eq:f_sim}---a scenario under which the normality assumption is actually satisfied---with $\sigma=1$ and the same $(\mu_1$, $\mu_2)$-combination. Every simulation is repeated 100 times, and the results over these 100 repetitions are reported in Table~\ref{tab:sim}.

\begin{table}[htp]
    \centering
    \begin{tabular}{cccc}
    \hline
    \multirow{3}{*}{Parameterization}  & \multirow{3}{*}{$|\mu_1-\mu_2|$}  
     &   \multicolumn{2}{c}{Mean Fraction of Incorrect Assignments} \\
    & & \multicolumn{2}{c}{(and its stardard error in brackets)} \\
                                           \cline{3-4} 
    & & $f(z)\sim$ Laplace \eqref{eq:f_sim} & $f(z)\sim$ Normal \eqref{eq:f_choice} \\
    \hline
        & 3 & 0.0385 (0.0019) & 0.0316 (0.0012) \\    
     QP & 2 & 0.0584 (0.0029) & 0.0465 (0.0019) \\
        & 1 & 0.1500 (0.0111) & 0.0990 (0.0058) \\
    \hline
        & 3 & 0.0307 (0.0022) & 0.0215 (0.0015) \\
   2LNN & 2 & 0.0437 (0.0031) & 0.0350 (0.0019) \\
        & 1 & 0.1233 (0.0098) & 0.0856 (0.0064) \\
    \hline
    \end{tabular}
    \caption{\label{tab:sim}Fraction of incorrect assignments made by our estimated boundary function when data are generated from Laplace versus Normal distributions, over 100 repeated simulations.}
\end{table}

Clearly, the performance of our algorithm deteriorates if the normality assumption \eqref{eq:f_choice} is violated, as one would expect, but the amount of deterioration is quite mild in all cases where the benchmark error (rightmost column of Table~\ref{tab:sim}) is relatively low (e.g., $< 0.05$). The deterioration only becomes more substantial when the benchmark error itself is already quite high. In our context, these would correspond to cases in which the values $z_1, \dots, z_n$ cannot be meaningfully separated into two groups; that is, there is not really a clear boundary  in the $(x,y)$-space beyond which changes in $z$ can be considered to have plateaued. Arguably, in such cases one shouldn't even attempt to apply our method in the first place. 

Table~\ref{tab:sim} also shows a clear advantage in parameterizing the boundary function as a 2LNN rather than as a QP, because here the true function $g(x,y)$ is sinusoidal, and no quadratic function can adequately approximate it. 

\bibliographystyle{apalike}
\bibliography{refs}
\end{document}